\documentstyle[aps,sprocl]{revtex}

\begin{document}

    \newcommand{\DSC}{D\hspace{-0.25cm}\slash_{\bot}}
    \newcommand{\DSP}{D\hspace{-0.25cm}\slash_{\|}}
    \newcommand{\DS}{D\hspace{-0.25cm}\slash}
    \newcommand{\DC}{D_{\bot}}
    \newcommand{\DSCX}{D\hspace{-0.20cm}\slash_{\bot}}
    \newcommand{\DSPX}{D\hspace{-0.20cm}\slash_{\|}}
    \newcommand{\DP}{D_{\|}}
    \newcommand{\QV}{Q_v^{+}}
    \newcommand{\QVB}{\bar{Q}_v^{+}}
    \newcommand{\QVP}{Q^{\prime +}_{v^{\prime}} }
    \newcommand{\QVBP}{\bar{Q}^{\prime +}_{v^{\prime}} }
    \newcommand{\QVHZ}{\hat{Q}^{+}_v}
    \newcommand{\QVHZB}{\bar{\hat{Q}}_v{\vspace{-0.3cm}\hspace{-0.2cm}{^{+}} } }
    \newcommand{\QVPHZB}{\bar{\hat{Q}}_{v^{\prime}}{\vspace{-0.3cm}\hspace{-0.2cm}{^{\prime +}}} }
    \newcommand{\QVPHFB}{\bar{\hat{Q}}_{v^{\prime}}{\vspace{-0.3cm}\hspace{-0.2cm}{^{\prime -}} } }
    \newcommand{\QVPHB}{\bar{\hat{Q}}_{v^{\prime}}{\vspace{-0.3cm}\hspace{-0.2cm}{^{\prime}} }   }
    \newcommand{\QVHF}{\hat{Q}^{-}_v}
    \newcommand{\QVHFB}{\bar{\hat{Q}}_v{\vspace{-0.3cm}\hspace{-0.2cm}{^{-}} }}
    \newcommand{\QVH}{\hat{Q}_v}
    \newcommand{\QVHB}{\bar{\hat{Q}}_v}
    \newcommand{\VS}{v\hspace{-0.2cm}\slash}
    \newcommand{\MQ}{m_{Q}}
    \newcommand{\MQP}{m_{Q^{\prime}}}
    \newcommand{\QVHPMB}{\bar{\hat{Q}}_v{\vspace{-0.3cm}\hspace{-0.2cm}{^{\pm}} }}
    \newcommand{\QVHMPB}{\bar{\hat{Q}}_v{\vspace{-0.3cm}\hspace{-0.2cm}{^{\mp}} }  }
    \newcommand{\QVHPM}{\hat{Q}^{\pm}_v}
    \newcommand{\QVHMP}{\hat{Q}^{\mp}_v}

\title{ \large \bf $V_{ub}$, $V_{cb}$ and Decay Constants in \\
Effective Field Theory of Heavy Quarks\footnote{Talk delivered by W.Y. Wang
at the International Conference on Flavor Physics (ICFP2001), May 31-June 6, Zhang-Jia-Jie, China.}}
\author{\small  W.Y. Wang, Y.L. Wu and Y.A. Yan}
\address{Institute of Theoretical Physics, Chinese Academy of Sciences, \\
Beijing 100080, China }
\maketitle

\begin{abstract}
A new framework of heavy quark effective field theory (HQEFT) is studied and compared
with the usual heavy quark effective theory (HQET). $|V_{ub}|$,
$|V_{cb}|$ and heavy meson decay constants are extracted in the new framework.
HQEFT can yield reasonable results for both exclusive
and inclusive decays.
\end{abstract}


\renewcommand{\theequation}%
 {\arabic{equation}}

\vspace{1cm}

The CKM matrix elements $V_{ub}$, $V_{cb}$ and the heavy meson decay constants
are of crutial importance in particle physics. They have been studied for a long time
in the full QCD and in effective theories, but their values are far from being determined precisely.
HQET is one of the most favorable effective theories in attacking heavy hadrons.
In the usual HQET, one generally decouples the quark and antiquark fields
and deals with only one of them independently. This only provides an approximate treatment for some
cases though it has been widely applied to various processes.

It has been noticed for a long time that the usual HQET does not always yield
consistent and reliable results. The decay constants of heavy mesons receive unexpectedly
large $1/m_Q$ order corrections \cite{neubprd1076,ballnpb}, which makes the calculation based on
heavy quark expansion inconsistent, in some sense.
For the lifetime ratios of $\Lambda_b$ baryon and B meson, the theoretical estimate in HQET
($\frac{\tau (\Lambda_b)}{\tau (B^0)} \ge 0.98 $ \cite{hk59}) appears to be too large so that it
conflicts with the world average of experimental measurements
($\frac{\tau (\Lambda_b)}{\tau (B^0)}=0.79\pm 0.05$ \cite{leplifetime}).

In the usual framework of HQET, $1/m_Q$ order corrections to the inclusive decay rate are absent
only when the rate is presented in terms of heavy quark mass $m_Q$ rather
than the heavy hadron mass $m_H$\cite{chay}. This seems to be conflict with the case in the
exclusive decays where the normalization of the transition matrix elements is given in terms
of heavy hadron mass\cite{luke}.
Such an inconsistency in the usual HQET may be the main reason that leads to the difficulty
in understanding the lifetime difference among the bottom hadrons.

Many processes have not yet been well investigated in HQET.
Since the usual HQET does not include the contributions of effective
antiquarks in the effective Lagrangian, it is not suitable for analyzing
the processes including quark-antiquark pair creation and annihilation. From this point of  view,
a complete Lagrangian with including both contributions of quarks and
antiquarks would be favorable. Transitions such as B decays into
excited charmed mesons and heavy-to-light decays have not yet been
studied to a satisfactory degree. The large uncertainties
require both more precise experimental measurements and
a reliable framework in theory.

The heavy quark spin-flavor symmetry
can be manifestly exhibited in effective theories.
This enables one to analyze different heavy hadrons and processes
by using the same set of universal functions, which are independent of the spin-flavor
of the heavy quarks.
In this way the heavy flavor physics becomes simpler for the reduction of independent parameters.
Though the HQET presents a compact framework, there are
possibilities of further reducing the number of independent wave functions in the HQEFT
as will be illustrated below (see also Refs.\cite{pub1,pub3}).

Strictly speaking, in quantum field theory particle and antiparticle decouple
completely only in the limit ${\MQ\rightarrow \infty}$.
To consider the finite quark mass correction precisely, it is necessary to include the
contributions from the components of the antiquark fields.
One can simply
extend the usual HQET to a heavy quark effective field theory with keeping
both effective quark and antiquark fields. This was first pointed out
by Y. L. Wu in Ref.\cite{ylw819}.

One may start from QCD Lagrangian:
\begin{equation}
  {\cal L}_{QCD}=
{\cal L}_{light} + {\cal L}_{heavy}=
{\cal L}_{light} + \bar{Q} (i\DS-m_Q) Q  .
\end{equation}
Firstly, denote the quantum field $ Q$ as
\begin{equation}
\label{QZQF}
  Q=Q^{+}+Q^{-}
\end{equation}
with $ Q^{+}$ and $ Q^{-}$ being the two solutions of the Dirac equation,
$  (i\DS-\MQ)Q^{\pm}=0 .$
We decompose the fields $Q^{\pm}$ into
\begin{equation}
   Q^{\pm}=\hat{Q}^{\pm}_v+R^{\pm}_v ,
\end{equation}
where
$      \hat{Q}^{\pm}_v\equiv \frac{1{\pm}v\hspace{-0.15cm}\slash}{2}Q^{\pm},
      R^{\pm}_v\equiv \frac{1{\mp}v\hspace{-0.15cm}\slash}{2}Q^{\pm}  $,
and introduce new field variables $Q_v$ and $\bar{Q}_v$ defined by
\begin{equation}
   Q_v=e^{iv\hspace{-0.15cm}\slash m_Q v\cdot x}\QVH ,\hspace{1.5cm}
       \bar{Q}_v=\QVHB e^{-i v\hspace{-0.15cm}\slash m_Q v \cdot x}.
\end{equation}
$ v^{\mu}$ can be an arbitrary four-vector satisfying $ v^2=1$.
Note that both $\QVH$ and $R_v$ contain the components of quark and antiquark
fields. $\hat{Q}^{\pm}_v$ are the `large components' and $R^{\pm}_v$ the `small components'.
With above definitions, an effective Lagrangian for the large components ($\QVH$ or
$Q_v$) can be easily derived by integrating out the small
components $R^{\pm}_v$:
\begin{equation}
\label{QCD2eff}
 {\cal L}_{heavy}   \to  {\cal L}_{Q,v}
      = {\cal L}^{++}_{Q,v}+{\cal L}^{--}_{Q,v}
          +{\cal L}^{+-}_{Q,v}+{\cal L}^{-+}_{Q,v}
\end{equation}
with
\begin{eqnarray}
    \label{eq:LZF}
   {\cal L}^{\pm \pm}_{Q,v}
      &=&   \bar{Q}^{\pm}_v [i\DSP +\frac{1}{2\MQ}i\DSC (1-\frac{i\DSP }{2\MQ})^{-1}
           i\DSC ] Q^{\pm}_v \equiv \bar{Q}^{\pm}_v A Q^{\pm}_v , \\
   {\cal L}^{\pm \mp}_{Q,v}
      &=&   e^{\pm 2im_Q v\cdot x} \bar{Q}^{\pm}_v [ -i\DSC+\frac{1}{4\MQ^2}
         (-i\stackrel{\hspace{-0.2cm}\leftarrow}{\DSC})
          (1-\frac{-i\stackrel{\hspace{-0.1cm}\leftarrow}{\DSP} }{2\MQ})
         ^{-1}i\DSC \nonumber \\
      &\times&   (1-\frac{i\DSP }{2\MQ})^{-1} i\DSC ]  Q^{\mp}_v
      \equiv e^{\pm 2im_Q v\cdot x} \bar{Q}^{\pm}_v B Q^{\mp}_v .
\end{eqnarray}

In the usual HQET, quark and antiquark fields are completely
decoupled, and the effective Lagrangian there includes only the $Q^+$ component
in Eq.(\ref{QZQF}). As a result, the effective Lagrangian in HQET contains only
${\cal L}^{++}_{Q,v}$ but not the other 3 parts in Eq.(\ref{QCD2eff}).

If further integrating out the antiquark fields (but not neglecting their
contributions), the effective Lagrangian can be
written in the following form in terms of only quark fields:
\begin{eqnarray}
\label{eq:23}
  {\cal L}_{eff}&=&  \QVB\{i\DSP+\frac{1}{2\MQ}i\DSC (1-\frac{i\DSP}
  {2\MQ})^{-1}i\DSC
   -[i\DSC+\frac{1}{2\MQ} i\DSC   \nonumber\\
&& \times (1-\frac{i\DSP}{2\MQ})^{-1}i\DSC
  \frac{1}{i\DSP}i\DSC]
 {[i\DSP-2\MQ-i\DSC\frac{1}{i\DSP}i\DSC]}^{-1} \nonumber\\
&&\times
 [i\DSC+\frac{1}{2\MQ}i\DSC \frac{1}{i\DSP}i\DSC
  (1-\frac{i\DSP}{2\MQ})^{-1}i\DSC]\}\QV .
\end{eqnarray}
  When the mass of a heavy quark is much larger than the QCD scale $\Lambda_{QCD}$,
this effective Lagrangian can be expanded in inverse power of
the heavy quark mass and be straightforwardly written as
$ {\cal L}_{eff}= {\cal L}^{(0)}_{eff}+{\cal L}^{(1/\MQ)}_{eff} $ with
\begin{eqnarray}
\label{eq:25}
     {\cal L}^{(0)}_{eff} & = &   \QVB(i\DSP)\QV ,  \\
    \label{eq:26}
      {\cal L}^{(1/\MQ)}_{eff} & = &   \frac{1}{\MQ}\QVB(i\DSC)^2\QV +
   \frac{1}{2\MQ^2}\QVB i\DSC (i\DSP)i\DSC \QV  \nonumber\\
          &  +& \frac{1}{4\MQ^2}\QVB (i\DSC)^2 \frac{1}{i\DSP}(i\DSC)^2 \QV
           +O(\frac{1}{\MQ^3}) ,
\end{eqnarray}

To show the contributions of antiquarks more clearly, the Lagrangian
can be written as:
 \begin{eqnarray}
    \label{eq:oth}
 {\cal L}&\to&  {\cal L}_{Q,v}\to {\cal L}_{eff}={\cal L}^{++}_{Q,v}+\tilde{\cal L}^{++}_{Q,v},\\
\label{eq:oth2}
 \tilde{\cal L}^{++}_{Q,v} & \equiv&    {\cal L}^{+-}_{Q,v}+{\cal L}^{-+}_{Q,v}
    +{\cal L}^{--}_{Q,v}  \nonumber\\
& =& \frac{1}{2m_Q}\QVB A \frac{1}{i\DSP} i\DSC
[1-\frac{i\DSP}{2m_Q}+\frac{1}{2m_Q}i\DSC \frac{1}{i\DSP}i\DSC ]^{-1}
  i\DSC \frac{1}{i\DSP} A \QV.
 \end{eqnarray}
$ \tilde{\cal L}^{++}_{Q,v}$ is the additional contributions to the part $ {\cal L}^{++}_{Q,v}$ that
is adopted in the usual HQET. This additional part may be regarded as the effective
potential of heavy quark due to the exchanges of virtual antiquarks.
When one imposes the on-shell condition $ AQ^+_v=0$, i.e.
 $ {\cal L}^{++}_{Q,v}=0$, the effective potential $ \tilde{\cal L}^{++}_{Q,v}$
also vanishes.
For an off-shell heavy quark in a hadron, $\tilde{\cal L}^{++}_{Q,v} \neq 0$,
and the leading order contribution of the effective potential is
$\tilde{\cal L}^{++}_{Q,v}|_{LO} =\QVB \frac{(i\DSCX)^2}{2m_Q} \QV $.
As is expected, antiquarks contribute from the order of $ 1/{\MQ}$.

The effective Lagrangian $ {\cal L}_{eff}$ automatically preserves the
velocity reparametrization invariance as well as Lorentz invariance
without the need of summing over the velocity. This can be understood easily by noticing
the fact
$  {\cal L}_{QCD}={\cal L}_{light}+{\cal L}_{Q,v}={\cal L}_{light}
     +{\cal L}_{Q,v'} . $

It is also noticed that Luke's theorem is a natural result of this HQEFT, without the need
of on-shell condition $ iv\cdot D\QV=0$ \cite{pub1}.
This result is nontrival and has its importance in explaining the bottom hadron
lifetime ratios in HQEFT.

Furthermore, fewer independent functions (parameters) are needed in
HQEFT than in HQET. For example, to characterize the $1/m_Q$ order corrections to
heavy-to-heavy transition matrix elements, HQET needs all together 8 functions,
while HQEFT needs only 3. And for the $1/m_Q$ order corrections to heavy meson decay
constants, HQET needs 4 parameters, while HQEFT needs only 2.
This simplification has been illustrated in detail in Refs.\cite{pub1,pub3}.

In the HQEFT, heavy meson decay constants can be written as \cite{pub3}
\begin{equation}
 f_M=\frac{F}{\sqrt{m_M}} \{ 1+\frac{1}{m_Q}(g_1+2d_M g_2)\}.
\end{equation}
$F$ is the leading order parameter, while $g_1$, $g_2$ characterize the $1/m_Q$ order corrections.
These have been calculated using sum rule methods with the results \cite{pub3}:
$\bar{\Lambda}=0.53\pm 0.08 $GeV, $ F=0.30\pm 0.06 \mbox{GeV}^{3/2}$,
$ g_1=0.54 \pm 0.12 $GeV, $ g_2 = -0.06\pm 0.02 $GeV.
Notice that $|g_1|/m_Q$ remains small enough so that the $1/m_Q$ expansion
in the new framework of HQEFT appears to be reliable. The scaling
law $f_M \sim F/\sqrt{m_M}$ is only slightly broken at the $1/m_Q$ order.

As a comparison, decay constants in the usual HQET is represented as
\begin{equation}
 f_M=\frac{F}{\sqrt{m_M}} (1-d_M \frac{\bar{\Lambda}}{6 m_Q}
   +\frac{G_1}{m_Q}+d_M \frac{G_2}{m_Q} ).
\end{equation}
with the $1/m_Q$ order corrections very large
($ G_1 \approx -2.0\mbox{GeV}$
in Ref.\cite{neubprd1076}, $ G_1 \approx -0.8\mbox{GeV} $ in Ref.\cite{ballnpb}).
So, in the HQET, the scaling law is seriously broken.

For the decay constants, we get in HQEFT when no QCD radiative corrections are
considered \cite{pub3}:
$  f_B(m_b)=159\pm 42$  MeV,
$   f_{B^{\ast}}(m_b)=166\pm 38$ MeV,
$   f_D(m_c)=251 \pm 99 $ MeV,
$ f_{D^{\ast}}(m_c)=293\pm 86 $MeV.
With two-loop QCD corrections, we get:
$  f_B(m_b)=196\pm 44 $ MeV,
$  f_{B^{\ast}}(m_b)=206\pm 39 $ MeV,
$ f_D(m_c)=298 \pm 109 $ MeV,
$  f_{D^{\ast}}(m_c)=354\pm 90 $ MeV.
These agree with results obtained from lattice simulations or other phenomenological
approaches \cite{othf}.

Semileptonic decays $B\to D^*(D)l\nu$ provide one of the main approaches to extract
the value of $|V_{cb}|$. The differential decay rates are:
\begin{eqnarray}
 \frac{d\Gamma(B\rightarrow D^{\ast}l\nu)}{d\omega} &=&
  \frac{G^2_F}{48\pi^3}(m_B-m_{D^{\ast}})^2
    m^3_{D^{\ast}}\sqrt{\omega^2-1}(\omega+1)^2 \nonumber\\
    &&\hspace{-1.5cm}   \times  [1+\frac{4\omega}{\omega+1}
    \frac{m^2_B-2\omega m_B m_{D^{\ast}}+m^2_{D^{\ast}}}
    {(m_B-m_{D^{\ast}})^2}]\vert V_{cb} \vert^2 {\cal F}^2(\omega) , \\
\frac{d\Gamma(B\rightarrow Dl\nu)}{d\omega} &=&
 \frac{G^2_F}{48\pi^3}(m_B+m_{D})^2
    m^3_{D}(\omega^2-1)^{3/2} \vert V_{cb} \vert^2 {\cal G}^2(\omega)
  \end{eqnarray}
 with $\omega=v\cdot v'$,
$    {\cal F}(\omega) = \eta_{A} h_{A_1}(\omega)$,
$    {\cal G}(\omega) = \eta_{V} [h_{+}(\omega)-\frac{m_B-m_D}{m_B+m_D}
      h_{-}(\omega)] $,
where
$ \vert V_{cb}\vert {\cal F}(1)=0.0352\pm 0.0026$,
$ \vert V_{cb}\vert {\cal G}(1)=0.0386\pm 0.0041$\cite{neu269};
$\eta_A=0.960\pm 0.007$, $\eta_V=1.022\pm 0.004$\nobreak\cite{ac}.

Due to the Luke's theorem, the factors $h_{A_1}$, $h_+$ and $h_-$ are protected from
$1/m_Q$ order corrections at the zero recoil point $\omega=1$. For this reason,
in the usual HQET, the $ B\rightarrow D^{\ast}l\nu$
decay is more favorable for the extraction of $ \vert V_{cb}\vert$, because its
decay rate receives no corrections of $ 1/m_Q$ order.
In the new framework of HQEFT, however,
$\Large  h_{-}(\omega)=0$ \cite{pub1}. This implies that both the differential
decay rates of channels
   $\Large  B\rightarrow Dl\nu$ and $\Large  B\rightarrow D^{\ast}l\nu$ receive
   no order $\Large  1/m_Q$ corrections. Therefore both channels can be used to extract
$|V_{cb}|$ reliably.

The HQEFT gives interesting relations
between meson masses and wave functions \cite{pub1}.
Those relations enable one to get the zero recoil values of HQEFT wave functions,
and furthermore to extract $|V_{cb}|$ from the meson masses.
We get $ \vert V_{cb} \vert = (3.78 \pm 0.28_{\mbox{exp}} \pm 0.18_{\mbox{th}})\times 10^{-2} $
 from $B\to D^* l\nu $ decay, and
$ \vert V_{cb} \vert = (3.82 \pm 0.41_{\mbox{exp}} \pm 0.28_{\mbox{th}} )\times 10^{-2}$
 from $B\to D l\nu$ decay \cite{pub1},
which are close to each other, and agree with the world average \cite{epjc2000}:
$ (3.67 \pm 0.23_{\mbox{exp}} \pm 0.18_{\mbox{th}})\times 10^{-2}$
  with using dispersion relation $\bf {\cal F}(\omega)$ parameterization,
and
$ ( 3.92 \pm 0.30_{\mbox{exp}} \pm 0.19_{\mbox{th}})\times 10^{-2} $
  with using linear $\bf {\cal F}(\omega)$ parameterization.

Semileptonic decays $B\to \pi (\rho) l\nu$ provide two of the most direct channels to
estimate $|V_{ub}|$.
One can parameterize the leading order transition matrix elements in the
effective theory of heavy quarks as \cite{gzmy,hly}
\begin{eqnarray}
 \langle \pi(p)|\bar{u} \Gamma \QV|B_v\rangle &=& -Tr[\pi(v,p)\Gamma {\cal M}(v)],\nonumber\\
\langle \rho(p,\epsilon^*)|\bar{u} \Gamma \QV|B_v\rangle
   & =& -i Tr[\Omega(v,p)\Gamma {\cal M}_v]   ,\nonumber
\end{eqnarray}
where
\begin{eqnarray}
  \pi(v,p)&=&  \gamma^5 [A(v\cdot p,\mu)+ {\hat{p}\hspace{-0.17cm}\slash}
 B(v\cdot p,\mu)], \nonumber\\
  \Omega(v,p)&=&    L_1(v\cdot p) {\epsilon\hspace{-0.17cm}\slash}^*
   +L_2( v\cdot p) (v\cdot \epsilon^*)
  + [L_3(v\cdot p)
   {\epsilon\hspace{-0.17cm}\slash}^* +L_4(v\cdot p) (v\cdot \epsilon^* )]
   {\hat{p}\hspace{-0.17cm}\slash}
\end{eqnarray}
with $\hat{p}^\mu=\frac{p^\mu}{v\cdot p}$.
A, B and $L_i(i=1,2,3,4)$ are the leading order wave functions characterizing the
heavy-to-light matrix elements in the
effective theory. In heavy-to-light decays, the HQS loses some of its predictive
power. However, HQS and relevant effective theories are still useful since they give
us relations between different channels. For example, $B\to \rho l\nu$ and
$D\to \rho l\nu$ are characterized by the same set of wave functions $L_i$.
We calculated these leading order wave functions using light cone sum rules.
Consider the $\pi$ distribution functions up to twist 4 and the $\rho$ distribution
functions up to twist 2,
we get $ |V_{ub}|=(3.4\pm 0.5_{\mbox{exp}} \pm 0.5_{\mbox{th}})
   \times 10^{-3} $ from $B\to \pi l\nu$ decay \cite{pub5},
and $ |V_{ub}|=(3.9\pm 0.6_{\mbox{exp}} \pm 0.7_{\mbox{th}} )\times 10^{-3} $
   from $ B\to \rho l\nu $ decay \cite{pub6}.
These leading order results agree well with those from full QCD calculations \cite{ar}:
$(3.9 \pm 0.6_{\mbox{exp}} \pm 0.6_{\mbox{th}} )\times 10^{-3}$ from $B\to\pi l \nu$ decay, or
$(3.4 \pm 0.6_{\mbox{exp}} \pm 0.5_{\mbox{th}} )\times 10^{-3}$ from $B\to\rho l \nu$ decay,
and are also compatible with the combined result from the analyses based on different
models and treatments on $B\to \pi (\rho) l \nu$ transitions\cite{cleo61}:
$|V_{ub}|=(3.25\pm 0.14^{+0.21}_{-0.29} \pm 0.55)\times 10^{-3}$.


Let us also take a look on the application of HQEFT on inclusive decays. The HQET
predictions on the lifetime defferences of bottom hadrons are \cite{hk59}:
$ {\tau (B^0_s) \over \tau (B^0)} \approx  1$,
${\tau (\Lambda_b) \over \tau (B^0)}  \geq  0.98 $,
while the world average values of experiments are \cite{leplifetime}:
$ {\tau (B^0_s) \over \tau (B^0)}=0.94 \pm 0.04 $,
${\tau (\Lambda_b) \over \tau (B^0)}=0.79 \pm 0.05 $.
This is a well-known contradiction in heavy flavor physics.
Calculations \cite{wwy2,wy} in HQEFT give
$ \frac{\tau (B^0_s)}{\tau (B^0)} = 0.96 \pm 0.06 $,
$ \frac{\tau (\Lambda_b)}{\tau (B^0)} =0.78\pm 0.05 $,
in good agreement with the experimental values.

Calculations in HQEFT on inclusive semileptonic decay rates also lead to
$|V_{cb}|$ and $|V_{ub}|$ values \cite{wwy2,wy}:
$ |V_{cb}|=(3.89 \pm 0.05_{\mbox{exp}}\pm 0.20_{\mbox{th}})\times 10^{-2} $,
$ |V_{ub}|=(3.48 \pm 0.62_{\mbox{exp}}\pm 0.11_{\mbox{th}})\times 10^{-3} $,
which agree well with those obtained from exclusive decays.

For the charm counting in $B^0$ decays, HQEFT predicts \cite{wwy2,wy}
$  n_c=1.19\pm 0.04 $,
which is smaller than predictions in HQET \cite{ebagan}:
$  n_c=1.24\pm 0.06 $ in $OS$ scheme, and
$ n_c=1.30\pm 0.07 $ in $\overline{MS}$ scheme.
So the value of $n_c$ obtained in HQEFT is, compared to the HQET prediction,
closer to the world average \cite{epjc}:
$  n_c=1.17 \pm 0.04 $.

In these applications of HQEFT into inclusive decays, the mass entering into the
inclusive decay rates is the
`dressed heavy quark' mass $\hat{m}^5_{Q}$ instead of the quark mass $ m^5_Q$,
where
\begin{equation}
\hat{m}_Q=m_Q + \bar{\Lambda} = m_{H_Q} (1+ O(1/m^2_{H_Q}))
\end{equation}
with $\bar{\Lambda}$ being the binding energy of the heavy hadron.
And therefore hadrons have different phase space effects in HQEFT.
These are the main reasons for the improvement of the new HQEFT with respect to the usual
HQET in studying inclusive decays.

As a summary, the HQEFT gives reasonable results for $|V_{ub}|$, $|V_{cb}|$
and heavy meson decay constants.
In all discussed applications in exclusive and inclusive decays,
the HQEFT proves to be consistent, reliable and simple. We found that the
antiquark components do contribute measurable effects, which in
turn supports the inclusion of both quark and antiquark contributions in the
effective Lagrangian. The HQEFT should be able to be applied to many other processes of
heavy hadrons, and further improve our understanding on the heavy flavor physics.

{\bf Acknowledgement} This work was supported in part by Chinese Academy of Sciences and by NSFC
under grant $\#$ 19625514.



\begin{thebibliography}{99}
\bibitem{neubprd1076} M. Neubert, Phys. Rev. D {\bf 46}, 1076 (1992).
\bibitem{ballnpb} P. Ball and V. M. Braun, Nucl. Phys. B {\bf 421}, 593 (1994).
\bibitem{hk59} H. Y. Cheng and K. C. Yang, Phys. Rev. D {\bf 59}, 014011 (1999).
\bibitem{leplifetime} For updated world averages of B hadron lifetimes and lifetime
    difference, see J. Alcaraz et. al. (LEP B Lifetime Group), http://wwwcn.cern.ch/
    ~claires/lepblife.html.
\bibitem{chay} J. Chay, H. Georgi, and B. Grinstein, Phys. Lett. {\bf B247}
          399 (1990).
\bibitem{luke} M. E. Luke, Phys. Lett. B {\bf 252}, 447 (1990).
\bibitem{ylw819} Y. L. Wu, Modern Phys. Lett. A {\bf 8}, 819 (1993).
\bibitem{pub1} W. Y. Wang, Y. L. Wu and Y. A. Yan, Int. J. Mod. Phys. A {\bf 15}, 1817 (2000).
\bibitem{pub3} W. Y. Wang and Y. L. Wu, Int. J. Mod. Phys. A {\bf 16}, 377 (2001).
\bibitem{othf} D. Becirevic, et. al., BUHEP-00-3, hep-lat/0002025;
L. Lellouch and C. D. Lin, CERN-TH/99-344, hep-ph/9912322;
C. Bernard et. al., Nucl. Phys. Proc. Suppl. {\bf 83}, 289 (2000);
S. Collins, et. al., Phys. Rev. D {\bf 60}, 074504 (1999).
\bibitem{neu269} M. Neubert, CERN-TH/98-2, hep-ph/9801269.
\bibitem{ac} A. Czarnecki, Phys. Rev. Lett. {\bf 22}, 4124 (1996).
\bibitem{epjc2000} Particle Data Group, Eur. Phys. J. C {\bf 15}, 2000.
\bibitem{gzmy} G. Burdman, Z. Ligeti, M. Neubert and Y. Nir, Phys. Rev. D {\bf 49},
2331 (1994).
\bibitem{hly} C. S. Huang, C. Liu and C. T. Yan, Phys.Rev. D {\bf 62}, 054019 (2000).
\bibitem{pub5} W. Y. Wang and Y. L. Wu, Phys. Lett. B {\bf 515}, 57 (2001).
\bibitem{pub6} W. Y. Wang and Y. L. Wu, Phys. Lett. B {\bf 519}, 219 (2001).
\bibitem{ar} A. Khodjamirian and R. R\"uckl, WUE-ITP-97-049, MPI-PhT/97-85, hep-ph/9801443.
\bibitem{cleo61} B. H. Behrens et. al.(CLEO Collab.), Phys. Rev. D {\bf 61}, 052001 (2000).
\bibitem{wwy2} Y. A. Yan, Y. L. Wu and W. Y. Wang, Int. J. Mod. Phys. A {\bf 15}, 2735 (2000).
\bibitem{wy} Y. L. Wu and Y. A. Yan, Int. J. Mod. Phys. A {\bf 16}, 285 (2001).
\bibitem{ebagan} E. Bagan, P. Ball, M. Braun and P. Gosdzinsky, Nucl. Phys. B {\bf 432},
3 (1994); Phys. Lett. B {\bf 342}, 362 (1995); Erratum, ibid. B {\bf 374}, 363 (1996);
E. Bagan, P. Ball, B. Fiol and P. Gosdzinsky, Phys. Lett. B {\bf 351}, 546 (1995).
\bibitem{epjc} Particle Data Group, Eur. Phys. J. C {\bf 3}, 1998.
\end{thebibliography}
\end{document}